\documentclass[showpacs,showkeys,11pt,
preprint,preprintnumbers,nofootinbib,
groupedaddress,superscriptaddress,amsmath,amssymb]{revtex4}
\usepackage{amsfonts}
\usepackage{graphics}
\usepackage{epsfig}
\usepackage{url}
\usepackage{multirow}
\usepackage{feynmp}
\newcommand {\be}{\begin{equation}}
\newcommand {\ee}{\end{equation}}
\newcommand {\ba}{\begin{eqnarray}}
\newcommand {\ea}{\end{eqnarray}}

\newcommand {\tanb}{$\tan\beta~$}
\newcommand {\ra}{\rightarrow}
\newcommand {\GeV}{\textnormal{GeV}~}

\begin{document}
\title{Charged Higgs Observability Through Associated Production With $W^{\pm}$ at a Muon Collider}
\author{M. Hashemi}
\email{hashemi_mj@shirazu.ac.ir}
\affiliation{Physics Department and Biruni Observatory, College of Sciences, Shiraz University, Shiraz 71454, Iran}

\begin{abstract}
The observability of a charged Higgs boson produced in association with a W boson at future muon colliders is studied. The analysis is performed within the MSSM framework. The charged Higgs is assumed to decay to $t\bar{b}$ and a fully hadronic final state is analyzed, i.e., $\mu^{+}\mu^{-} \ra H^{\pm}W^{\mp} \ra t\bar{b}W^{-} \ra W^{+}b\bar{b}W^{-} \ra jjjjb\bar{b}$. The main background is $t\bar{t}$ production in fully hadronic final state which is an irreducible background with very similar kinematic features. It is shown that although the discovery potential is almost the same for a charged Higgs mass in the range $200 ~\textnormal{GeV} < m_{H^{\pm}} < 400 ~ \textnormal{GeV}$, the signal significance is about $1\sigma$ for \tanb = 50 at integrated luminosity of $50~ fb^{-1}$.
The signal rate is well above that at $e^{+}e^{-}$ linear colliders with the same center of mass energy and enough data (O(1 $ab^{-1}$)) will provide the same discovery potential for all heavy charged Higgs masses up to $m_{H^{\pm}} \sim 400  ~\textnormal{GeV}$, however, the muon collider cannot add anything to the LHC findings.
\end{abstract}

\maketitle

\section{Introduction}
The Standard Model of particle physics (SM) contains one complex Higgs doublet. After electroweak symmetry breaking through the Higgs mechanism \cite{Higgs1,Higgs2,Higgs3,Higgs4,Higgs5}, a single neutral Higgs boson is produced. There are theoretical arguments against the standard model which motivate the idea that the standard model may not be the ultimate complete theory of elementary particles. One of such cases is the divergence of the Higgs boson mass when radiative corrections are included. The divergence quadratically depends on the ultraviolet momentum cut-off of the loop integrals which can be regarded as the ultimate energy scale of the theory, up to which no deviation appears from a theory or a new structure altering the high energy behavior. Obviously at the Planck scale $M_{P} \sim 10^{18}~ \textnormal{GeV}$, quantum gravitational effects require a new structure, however, there is still no hint of new phenomena beyond the SM at the experimental frontier about the TeV scale. Therefore the momentum cut-off is usually taken as $M_{P}$ which is $\sim 10^{16}$ orders of magnitude different from electroweak scale. With this assumption radiative corrections result in very large and un-natural values for the Higgs boson mass. One of the elegant solutions to this problem is the idea of supersymmetry which reduces the above divergence and retains the Higgs boson mass at a finite scale \cite{Martin}. 
In supersymmetric theories there is a non-minimal Higgs sector. For example, the Minimal Supersymmetric Standard Model (MSSM) is a Two Higgs Doublet Model (2HDM) \cite{2hdm} which comprises two Higgs complex doublets. After electroweak symmetry breaking, five physical Higgs bosons are produced: two neutral CP-even Higgs bosons ($h^{0},H^{0}$), one neutral CP-odd Higgs boson ($A^{0}$) and two CP-even charged Higgs bosons ($H^{\pm}$). Observation of more than one neutral Higgs boson may be an indication of 2HDM, however, an equivalent signature of these models is the charged Higgs boson which possesses different phenomenological characteristics.\\ 
Current experimental searches have set limits on the mass of this particle as a function of \tanb which is the ratio of vacuum expectation values of the two Higgs fields used to make the two Higgs doublets. The LEP II experiment has excluded charged Higgs mass lighter than $89~ \textnormal{GeV}$ for all \tanb assumptions \cite{lepexclusion1}. The indirect limits are however stronger, excluding charged Higgs masses lighter than $125~ \textnormal{GeV}$ \cite{lepexclusion2}. \\
The above results are followed by Tevatron searches reported in \cite{d01,d02,d03,d04} by the D0 Collaboration and \cite{cdf1,cdf2,cdf3,cdf4} by CDF Collaboration. The overall result is $2 <$ tan$\beta < 30$ for $m(H^{\pm}) > 80 $ GeV while more $\tan \beta$ values are available for higher charged Higgs masses.\\
The B-Physics constraints are also imposed on the charged Higgs mass as indirect limits. This is due to the fact that including a charged Higgs in B-meson decay diagrams may result in different decay rates and can be verified experimentally. The strongest such limits comes from a study of the $b\ra s\gamma$ transition process using CLEO data which excludes a charged Higgs mass below 295 GeV at 95 $\%$ C.L. in 2HDM Type II with \tanb higher than 2 \cite{B1}. On this direction two points should be considered. First, such constraints are model parameter dependent as it has been shown in \cite{B2} that in a general 2HDM with complex Higgs-fermion coupling, the charged Higgs mass can be as low as 100 GeV. Second, one should remember that these constraints belong to general 2HDMs which are not necessarily assumed to be supersymmetric. The MSSM framework studied in this paper belongs to a 2HDM type II and incorrporates supersymmetry. Allowing existence of sparticles may change the contraints obtained in B-Physics studies, and in general, a translation of those constraints frmo B-Physisc studies to the case of MSSM may not be trivial. \\ 
The most recent results on the charged Higgs searches come from ATLAS and CMS experiments at LHC \cite{atlasdirect,cmsdirect}. They use integrated luminosities of 4.6 and 2.3 $fb^{-1}$ respectively and both indicate that a light charged Higgs up to about $m_{H^{\pm}}\sim 140 ~\textnormal{GeV}$ is excluded with \tanb $>~ 10$. \\
The above searches indicate that a heavy charged Higgs may be in fact difficult to observe due to the hadronic environment of LHC collisions, large final state particle multiplicity and the large background cross sections. In addition the signal cross section decreases at heavy mass region leaving few events for some charged Higgs masses. This situation motivates a linear collider with leptonic input beams. Currently several proposals for constructing linear lepton colliders are under consideration and feasibility studies are on going \cite{ilc,rdr,clic,clic_cdr}. This is contrary to the last decade which was mainly dedicated to proton collisions in Fermilab and CERN. In lepton colliders, the incoming leptons (electrons or muons) have a sharper energy distribution than that at a hadron collider, collision events are usually cleaner with less hadronic activities and isolated leptons can be more easily identified and used for the signal selection. \\
\section{Charged Higgs at Future Linear Colliders}
There has been an extensive work on estimating future linear colliders potential for the discovery of Higgs bosons within MSSM or a more general 2HDM framework. At $e^{+}e^{-}$ linear colliders, the center of mass energy is expected to be 0.5 TeV at ILC \cite{ilc,rdr} and 0.5 to 3 TeV at CLIC \cite{clic,clic_cdr}. The target integrated luminosity for many analyses has been 500 $fb^{-1}$, although higher values are also expected when the migration from the low energy phase to the high energy phase is performed. Such colliders are expected to be complementary to LHC in finding new particles, testing new phenomena and precision measurements of physical parameters. Since a charged Higgs particle may be an indication of beyond SM, special care has been taken in searches for this particle. Analyses of this kind discuss the possibility of observing MSSM charged Higgs particles in case LHC fails to do so, or confirmation of signals observed at LHC.\\
The LHC potential for a charged Higgs observation depends strongly on the mass of this particle and \tanb. The main production process for a light charged Higgs at LHC would be a top pair production followed by the top quark decay to charged Higgs which decays subsequently to $\tau\nu$ \cite{CMSLCH}. This channel has sufficient cross section and will soon cover the accessible region of the parameter space. The result may be either a signal of a light charged Higgs or exclusion of the parameter space due to the missing particle's signal. The heavy charged Higgs at LHC may be produced through $gg\ra t\bar{b}H^{-}$ and $gb \ra tH^{-}$ described in \cite{ggtbh1,ggtbh2}, followed by its decay to $\tau\nu$ \cite{CMSHCH} or $t\bar{b}$ \cite{LowetteCH}. The cross section of this process decreases with increasing the charged Higgs mass, however, before a Muon Collider concept comes to reality, LHC may accumulate enough data for a heavy chargged Higgs observation through its low luminosity, high luminosity and probably Super-LHC phases. Therefore the role of a linear collider would be confirmatioon of LHC results and stepping forward towards precision measurements of the MSSM Higgs sector parameters. At $e^{+}e^{-}$ linear colliders, the charged Higgs can be produced in pair or singly. The pair production $e^{+}e^{-}\ra H^{+}H^{-}$ is limited to $m_{H^{\pm}}\leq \sqrt{s}/2$ \cite{pair}, however, the single charged Higgs production can probe areas of the parameter space not accessible by the pair production process \cite{sch1,sch2,sch3}. Analyses of the single heavy charged Higgs production through $e^{+}e^{-}\ra \tau\bar{\nu} H^{+}$ has proved that including off-shell effects leads to promising results \cite{sch6,sch7}. \\
The $W^{\pm}H^{\mp}$ process has also been studied extensively. Theoretical and phenomenological studies of this process with LHC type of collisions can be found in \cite{24,25,26,27,28,29,30,31,32,33,34,35,36,37}. A generator of this process is also available for proton-proton collisions \cite{pybbwh}. In a study reported in \cite{myWH} it was shown that using this channel alone, leads to better results than the current Tevatron limits on the charged Higgs searches. \\
At linear $e^{+}e^{-}$ colliders, there has been numerous studies of this channel. The theoretical background comes from \cite{WH1,WH2} where the cross section of this channel was calculated followed by studies of possible enhancement of the cross section by including quark and Higgs-loop effects in \cite{WH3}. The small cross section of this process has led to the conclusion that only few events of this kind may be observed at $e^{+}e^{-}$ linear colliders \cite{WH4,WH5} and therefore it seems hopeless at $e^{+}e^{-}$ colliders. However a study of this channel showed that it may be detectable at high \tanb values at a muon collider \cite{mumuWH} due to the Higgs-lepton Yukawa coupling enhancement by a factor of $m_{\mu}/m_{e}$ when using a $\mu^{+}\mu^{-}$ collider instead of an $e^{+}e^{-}$ collider. As indicated in \cite{mumuWH} the cross section of this process is almost independent of the charged Higgs mass and increases with increasing \tanb. The higher cross section of this channel at muon colliders compared to $e^{+}e^{-}$ colliders, is a motivation for this study. Therefore the aim of this analysis is to investigate the possibility of observing a charged Higgs boson in the heavy mass region, i.e., $200 ~\textnormal{GeV} < m_{H^{\pm}} < 400 ~\textnormal{GeV}$ at future muon colliders operating at a center of mass energy of 500 GeV. In addition to MSSM scenario, a similar study of this channel in \cite{mumuWH2} showed that in a general 2HDM with CP-violating terms, the signal cross section could be much higher than that in MSSM, however, we rely on the standard CP-conserving MSSM in this analysis.\\ 
\section{A Muon Collider and a Brief Review of its Physics Potential}
The muon collider concept has attracted attention in the High Energy Physics community for several reasons among which one can mention the less Synchrotron radiation in the circular path of the collider compared to an $e^{+}e^{-}$ collider, less beam smearing resulting in a very good beam energy resolution suitable for precision measurements and opportunity to perform precision measurements in the s-channel collisions which can be used, e.g.,  for Higgs boson mass measurements through channels like $h\ra b\bar{b}$. \\
There has been a proposal for a muon collider at the Fermilab. One of the original documents and a more recent one can be found in \cite{m1} and \cite{m2}. Such a collider is foreseen to be operating at two stages of the center of mass energy and luminosity. The first stage is a First Muon Collider (FMC) with a center of mass energy of 500 GeV, collecting 20 $fb^{-1}$ integrated luminosity per year, and a Next Muon Collider (NMC) operating at a center of mass energy of 4 TeV, with a capability of collecting 1000 $fb^{-1}$ integrated luminosity per year is foreseen for the second phase. \\
The physics potential of the muon collider, both FMC and NMC, has been studied in various subjects. A theoretical introduction to the physics of muon colliders is presented in \cite{JEllis}. A comprehensive overview of the physics potential of muon colliders can also be found in \cite{m3,m4,m5}. The $s-$channel Higgs physics has been studied in \cite{m6,m7,m8,m9}. These reports rely on the fact that the $s-$channel Higgs boson production cross section is enhanced at a muon collider by a factor of ${(m_{\mu}/m_{e})}^2\simeq 4\times 10^{4}$ and therefore address questions of observability of an SM or SM-like MSSM Higgs boson in the $s-$channel production and the possibility of distinguishing them as well as approaches to precision measurements of the Higgs boson mass as a function of the energy resolution of the beam which is much less than that at an $e^{+}e^{-}$ collider.\\
In this work, the focus is on $\mu^{+}\mu^{-} \ra H^{\pm}W^{\mp}$, i.e., a charged Higgs boson produced in MSSM which belongs to 2HDM Type II. The Higgs physics of muon colliders in 2HDM Type II has been studied theoretically in \cite{m10}. As is seen in \cite{m10} and \cite{mumuWH} the cross section of this channel is few tens of femtobarn at muon colliders and is almost flat with respect to the charged Higgs mass. This feature may raise the hope to observe a charged Higgs with equal chance in the range $200 ~\textnormal{GeV} < m_{H^{\pm}} < 400~ \textnormal{GeV}$. In the following we investigate this point in a quantitative way.
\section{Signal and Background Processes and Simulation Tools}
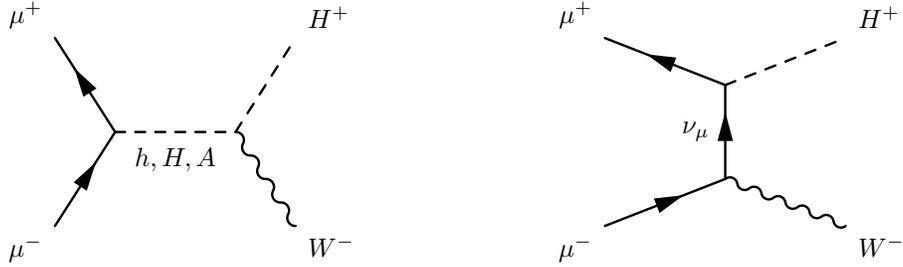
\begin{figure}
\begin{minipage}[t]{0.4\linewidth}
\centering
\unitlength=1mm
\begin{fmffile}{schannel}
\begin{fmfgraph*}(40,25)
\fmfleft{i1,i2}
\fmfright{o1,o2}
\fmflabel{$\mu^-$}{i1}
\fmflabel{$\mu^+$}{i2}
\fmflabel{$H^+$}{o2}
\fmflabel{$W^-$}{o1}
\fmf{fermion}{i1,v1,i2}
\fmf{dashes}{v2,o2}
\fmf{photon}{v2,o1}
\fmf{dashes,label=$h,,H,,A$}{v1,v2}
\end{fmfgraph*}
\end{fmffile}
\end{minipage}
\hspace{0.5cm}
\begin{minipage}[t]{0.4\linewidth}
\centering
\unitlength=1mm
\begin{fmffile}{tchannel}
\begin{fmfgraph*}(40,25)
\fmfleft{i1,i2}
\fmfright{o1,o2}
\fmflabel{$\mu^-$}{i1}
\fmflabel{$\mu^+$}{i2}
\fmflabel{$H^+$}{o2}
\fmflabel{$W^-$}{o1}
\fmf{fermion}{i1,v1}
\fmf{fermion}{v2,i2}
\fmf{fermion,label=$\nu_{\mu}$}{v1,v2}
\fmf{dashes}{v2,o2}
\fmf{photon}{v1,o1}
\end{fmfgraph*}
\end{fmffile}
\end{minipage}
\caption{The $s-$channel (left) and $t-$channel (right) diagrams involved in the signal process. \label{diagrams}}
\end{figure}

The signal to study is a charged Higgs produced in association with a W boson in $\mu^{+}\mu^{-}$ collisions, i.e., 
\be
\mu^{+}\mu^{-} \ra H^{\pm}W^{\mp} \ra t\bar{b}W^{-} \ra W^{+}b\bar{b}W^{-} \ra jjjjb\bar{b}.
\ee
In writing the above chain, a charged Higgs decay to $t\bar{b}$ pair and W boson decay to quarks (initiating jets) have been assumed. Therefore a fully hadronic final state is analyzed to enable us to perform a charged Higgs candidate mass reconstruction. The above process can proceed in both $s-$channel and $t-$channel diagrams as shown in Fig. \ref{diagrams}.\\
The $m_{h}-max$ benchmark scenario is used with $M_{2}=200$ GeV, $M_{\tilde{g}}=800$ GeV, $\mu=200$ GeV and $M_{SUSY}=1$ TeV. This setting corresponds to a light SM-like neutral Higgs ($h$) and four heavy Higgs bosons with almost degenerate masses. Table \ref{hmasses} shows the Higgs boson masses for \tanb = 50. The resulting values slightly depend on \tanb, e.g., with $m_{H^{\pm}}=400$ GeV, $m_{h}=128~(130)$ GeV for \tanb$=10~(50)$. Therefore the scenario is in agreement with LHC observation of a light neutral Higgs boson \cite{atlas125,cms125}.
\begin{table}
\begin{center}
\begin{tabular}{|c|c|c|c|}
\hline
$m_{H^{\pm}}$ [GeV] & $m_{A}$ [GeV] & $m_H$ [GeV] & $m_{h}$ [GeV]\\ 
\hline
200 &175 & 175 & 130\\ 
\hline
300 & 285 & 285 & 130\\
\hline
400 & 389 & 389 & 130\\
\hline
\end{tabular}
\caption{The Higgs boson masses used in the simulation of signal events. The values have been obtained with \tanb=50 \label{hmasses}}
\end{center}
\end{table}
The main background is $t\bar{t}$ process with $t\ra W^{+}b$ and W boson hadronic decay, thus producing the same final state. Since the final states are the same with the same particle type and multiplicity, reducing the $t\bar{t}$ background is a challenge. In fact what is observed is that the signal and background kinematics and therefore selection efficiencies are very similar. The signal cross section with \tanb = 50 and $m_{H^{\pm}}=200$ \GeV is about 25 $fb$ while the $t\bar{t}$ background has a cross section of 560 $fb$. Therefore the initial ratio of signal to $t\bar{t}$ background is $\sim 4 \%$. The $W^{+}W^{-}$ background is not expected to mimic the signal as it is expected to be suppressed by the double b-tagging (the requirement of having two b-jets in the event). Two other background samples, i.e., QCD background and $W^+W^-jj$ were also analyzed. The QCD events, i.e., $\mu^+\mu^- \ra j j$ including b-jets in the final state, have a cross section of 2.6 $pb$, however, they did not survive event selection. The $WWjj$ background was seperated into two samples of $W^+W^-jj$ with light jets in the final state and $W^+W^-b\bar{b}$ with two b-jets in the final state. The first sample includes only light jets and its contribution to the signal should be negligible as the efficiency of mistagging two light jets as b-jets is very small. These events should be simulated and analyzed in detail, however, they correspond to a large number of Feynman diagrams ($\sim$ 500). The $W^+W^-b\bar{b}$ with a cross section of $\sim 260$ $fb$ was instead analyzed and included in the final results.\\
For the simulation of the signal and the background, CompHEP package is used \cite{comphep,comphep2}. Events are written in LHEF format \cite{lhef} and passed to PYTHIA 8.1.53 \cite{pythia} for the fragmentation and hadronization and particle decay showers. For the SUSY spectrum and decay simulation the SUSY-HIT package \cite{susyhit} is used. This package contains HDECAY \cite{hdecay} built-in for the charged Higgs decay calculation. For the calculation of the particle spectrum, SUSY-HIT uses the renormalization group evolution program SuSpect \cite{suspect}. The output including the particles mass spectra and decays is written in SLHA format \cite{slha} and used by PYTHIA for event generation. \\
The output of the PYTHIA is translated to HEPMC 2.05.01 format \cite{hepmc} for event analysis. Having generated events, the event jets are reconstructed using FASTJET 2.4.1 \cite{fastjet}. The anti-kt algorithm \cite{antikt} and a cone size of 0.4 and the ET recombination scheme are used for the jet reconstruction. Finally when events are generated, kinematic distributions are visualized and analyzed using ROOT 5.30 \cite{root}.\\
\section{Signal and Background Cross sections}
\begin{figure}
\begin{center}
\includegraphics[width=0.7\textwidth]{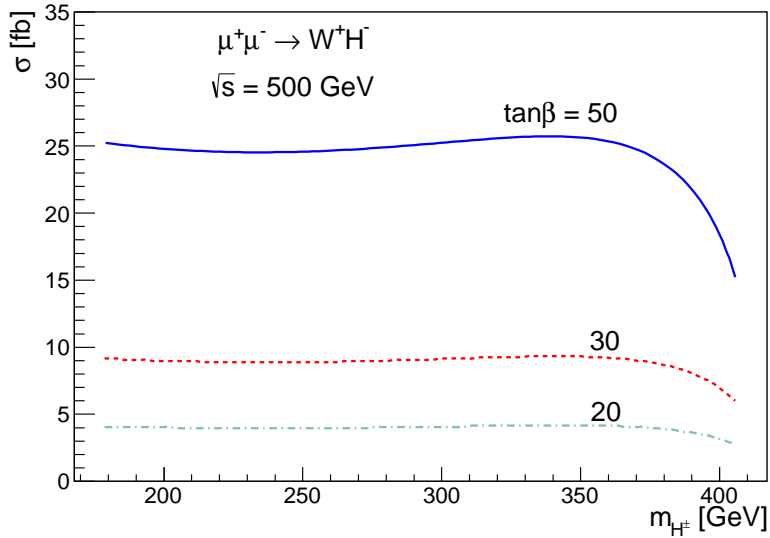}
\end{center}
\caption{The signal cross section as a function of the charged Higgs mass with various \tanb values.}
\label{sxsec}
\end{figure}

\begin{figure}
\begin{center}
\includegraphics[width=0.7\textwidth]{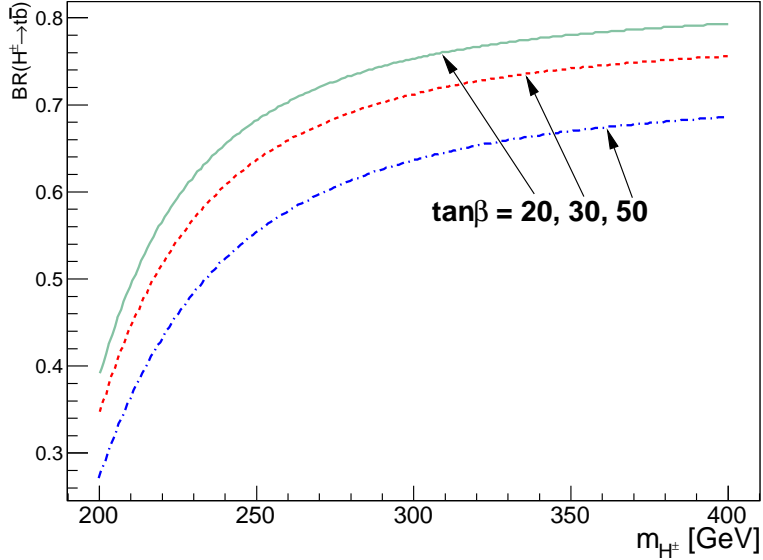}
\end{center}
\caption{The charged Higgs branching ratio of decay to $t\bar{b}$ with various \tanb values.}
\label{BR}
\end{figure}

The signal cross section was calculated using CompHEP. Figure \ref{sxsec} shows the signal cross sections as a function of the charged Higgs mass for various values of \tanb.
The cross sections in Fig. \ref{sxsec} should be multiplied by the charged Higgs branching ratio of decay to $t\bar{b}$ to get the right final state. The branching ratios obtained with HDECAY are shown in Fig. \ref{BR} for different \tanb values. 
The $t\bar{t}$ background total cross section is 560 $fb$ as stated before. In calculating $\sigma \times$BR for both signal and the background, BR$(t\ra W^{+}b)~=~1$ and BR$(W^{+}\ra jj)~=~0.68$ has been assumed.   
\section{Event Selection and Analysis Strategy}
\begin{figure}
\begin{center}
\includegraphics[width=0.7\textwidth]{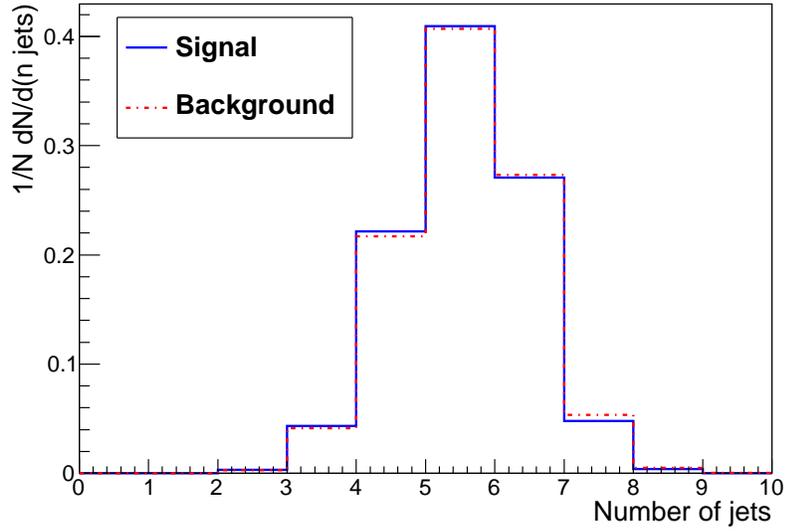}
\end{center}
\caption{The total jet multiplicity in signal and background events.}
\label{jetmul}
\end{figure}

\begin{figure}
\begin{center}
\includegraphics[width=0.7\textwidth]{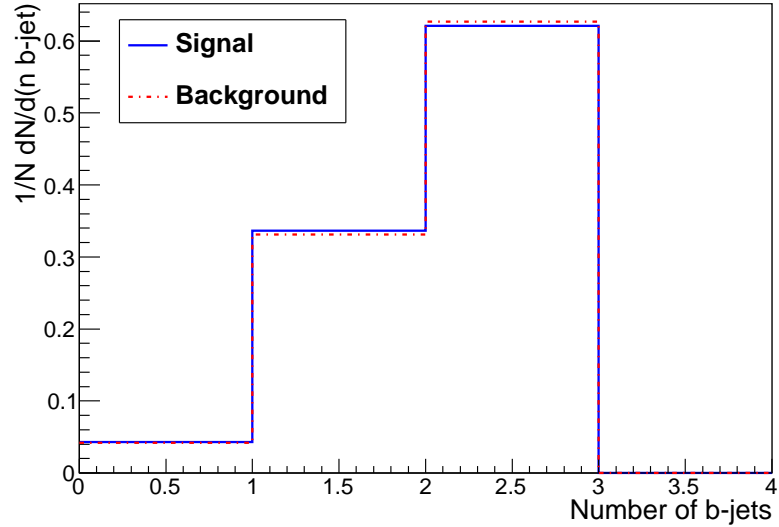}
\end{center}
\caption{The b-jet multiplicity in signal and background events.}
\label{bjetmul}
\end{figure}

\begin{figure}
\begin{center}
\includegraphics[width=0.7\textwidth]{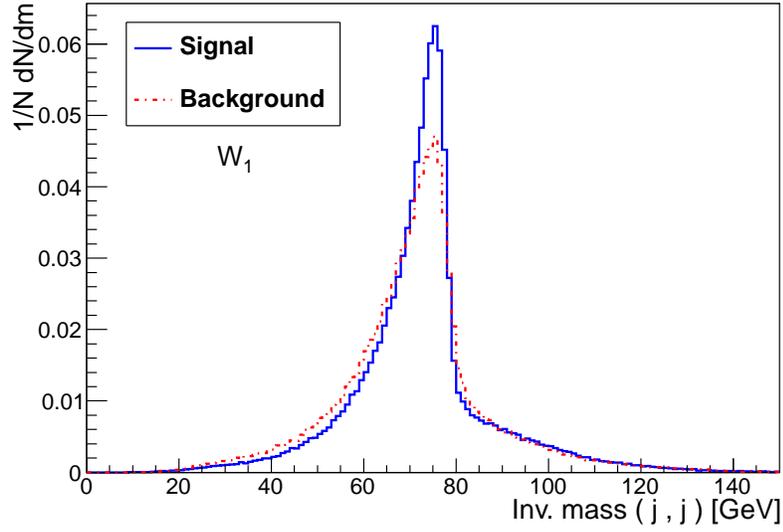}
\end{center}
\caption{The two jet invariant mass distribution as the W boson candidate in signal and background events.}
\label{w1}
\end{figure}

\begin{figure}
\begin{center}
\includegraphics[width=0.7\textwidth]{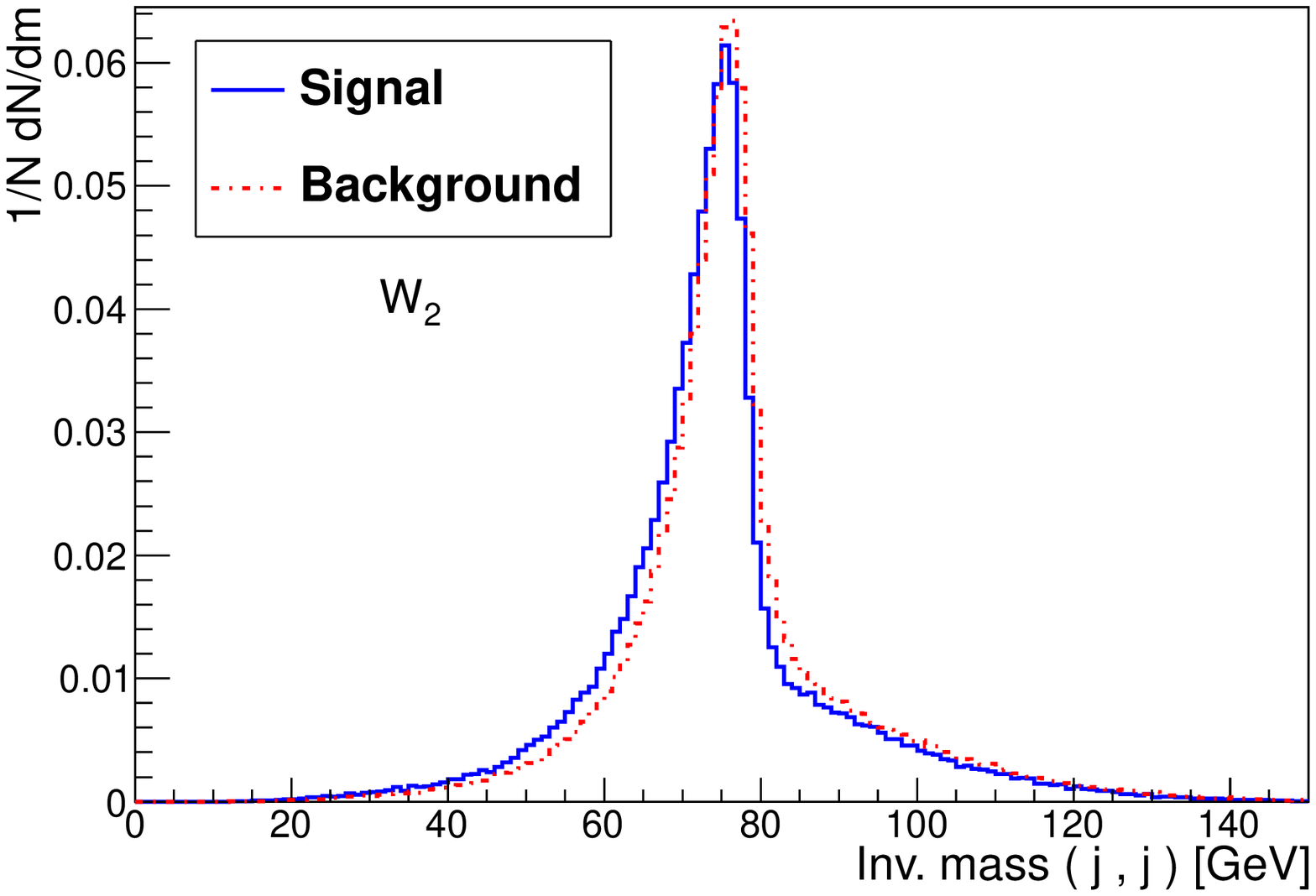}
\end{center}
\caption{The two jet invariant mass distribution as the W boson candidate from the top quark decay in signal and background events.}
\label{w2}
\end{figure}

\begin{figure}
\begin{center}
\includegraphics[width=0.7\textwidth]{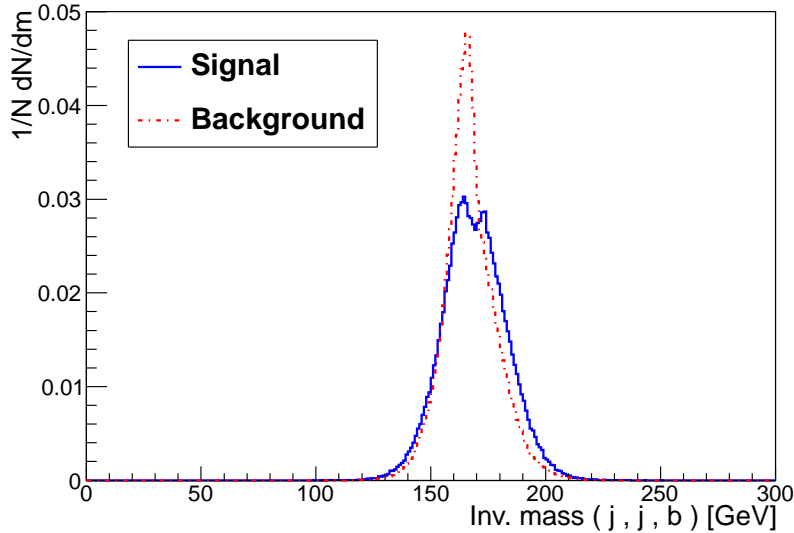}
\end{center}
\caption{The three jet invariant mass distribution as the top quark candidate in signal and background events.}
\label{top}
\end{figure}

\begin{figure}
\begin{center}
\includegraphics[width=0.7\textwidth]{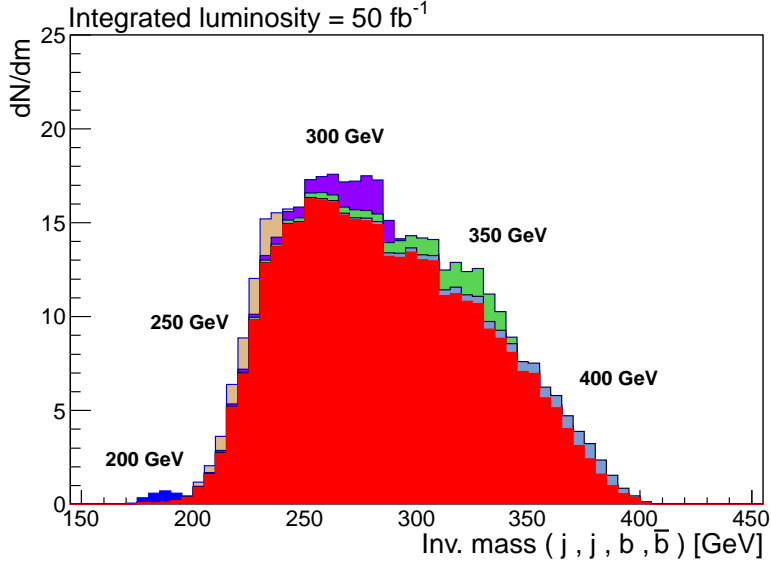}
\end{center}
\caption{The charged Higgs candidates on top of the $t\bar{t}$ background scaled to the expected number of events at an integrated luminosity of 50 $fb^{-1}$. Each signal distribution is lying independently on top of the background and the numbers in GeV denote the charged Higgs mass used in the simulation of the corresponding signal. The $WWb\bar{b}$ is not included due to the low statistics available.}
\label{signals}
\end{figure}

\begin{figure}
\begin{center}
\includegraphics[width=0.7\textwidth]{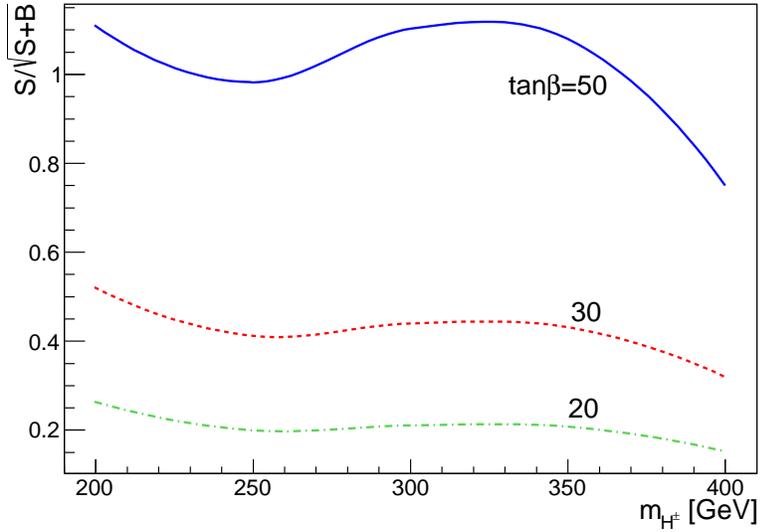}
\end{center}
\caption{The signal significance after the cut on the charged Higgs invariant mass. The $t\bar{t}$ and $WWb\bar{b}$ are included as the background.}
\label{sigma}
\end{figure}

The analysis of the signal and the background samples is started on an event-by-event basis, trying to reconstruct the jets, applying a cut on the number of jets, and preforming a W, top quark and charged Higgs reconstruction. The details of the analysis is presented as follows.\\
First a jet reconstruction is applied on the event, selecting only jets satisfying the following requirements:
\be
E_{T,\textnormal{jet}}~>~20~\GeV, ~~~~|\eta_{\textnormal{jet}}|~<~2.5
\ee
where $\eta=-\textnormal{ln}\tan(\theta/2)$ is the pseudorapidity. An event has to have 6 jets passing the above requirement, two of which are b-tagged. The b-tagging is emulated by a jet-quark matching algorithm, which calculates the spatial distance between the jet and a quark in terms of $\Delta R=\sqrt{{(\Delta\eta)}^2+{(\Delta\phi)}^2}$ where $\phi$ is the azimuthal angle. If $\Delta R$ between the jet and a b-quark in the event is less than 0.2, the jet is flagged as a b-jet otherwise it is a light jet (a jet from a light quark). Since the b-tagging algorithm efficiency in a real environment is not known, no b-tagging efficiency is applied here. Figures \ref{jetmul} and \ref{bjetmul} show the total jet and b-jet multiplicity distributions in signal and background events. All signal distributions are shown with $m_{H^{\pm}}=300$ \GeV hereafter. 
Selected events with 6 jets two of which are b-jets, are used in the next step for a W and top quark reconstruction. A $\chi^{2}$ minimization is used for finding the best jet combination in the event giving the closest possible values of reconstructed invariant masses to the nominal values of the W boson mass ($m_{W}=80.4$ GeV) and the top quark mass ($m_{t}=$173 GeV).
The $\chi^{2}$ definition is as in Eq. \ref{chi2}
\be
\chi^{2}={\left(\frac{m_{(j_i,j_j)}-m_{W}}{\sigma_W}\right)}^2+{\left( \frac{m_{(j_k,j_l)}-m_{W}}{\sigma_W}\right)}^2 +{\left( \frac{m_{(j_k,j_l,j_m)}-m_{t}}{\sigma_t}\right)}^2 
\label{chi2}
\ee
where the loop runs over the indices $i,\cdots, m$ provided that no pair of them are the same. The weights in the denominators are taken to be $\sigma_W=15$ \GeV and $\sigma_t=20$ \GeV. These are related to the spread of the invariant mass distributions of the W's and top quark. The reconstructed W's (the invariant mass distribution of the two jet pairs making the W candidates) are shown in Figs. \ref{w1} and \ref{w2}. The numbering scheme for signal events is $\mu^+\mu^- \ra W_{1}H^{\pm} \ra W_{1}t\bar{b} \ra W_{1}W_{2}b\bar{b}$. For the background, this means that the best reconstructed top quark with the closest invariant mass to the top quark mass is selected and the W boson associated to that (its decay product) is taken as $W_2$. Therefore the other W boson in the event is $W_{1}$. The reconstructed top quark is shown in Fig. \ref{top}. Since the signal statistics used in the analysis is large and the $t\bar{t}$ background distribution of the top quark candidate mass has a smooth distribution and a single peak, the two kinks in the signal distribution could be due to the right and wrong combination of jets in the top quark reconstruction.   
At this point, five jets of the event have been used for the W and top mass reconstruction. A mass window cut is now applied before proceeding to the charged Higgs mass reconstruction. Since the jet energies and their directions are subject to uncertainties due to the jet reconstruction algorithm, the reconstructed W and top mass distributions are slightly different from what is expected from the true values used in the simulation. This is a known feature and is expected to be resolved by the jet correction algorithms which are detector dependent and their study is beyond the scope of this analysis. The mass windows are thus applied on the observed distributions according to Eq. \ref{masswindow}. The superscript (rec.) denotes the reconstructed value of the particle mass. 
\be
60~ <~ m^{\textnormal{rec.}}_{W}~ <~ 90 ~\GeV,~~~~150~ <~ m^{\textnormal{rec.}}_{t}~ <~ 200 ~\GeV
\label{masswindow}
\ee
The reconstructed top quark is used together with the remaining b-jet in the event to make a charged Higgs candidate. The reconstructed charged Higgs candidates invariant masses are shown in Fig. \ref{signals}.
The number of events shown on Fig. \ref{signals} are correctly scaled to what is expected at an integrated luminosity of 50 $fb^{-1}$. To obtain those numbers, the total selection efficiencies have been used. Table \ref{seleff} shows the relative selection efficiencies for signal and background samples. As is seen, the signal and background selection efficiencies are very similar due to the very similar kinematics of events. In the best case, with $m_{H^{\pm}}=300$ GeV, a signal selection efficiency of 6$\%$ is obtained which is twice that of the $t\bar{t}$ background (3$\%$). In order to further increase the signal to background ratio, a mass window is applied on the charged Higgs candidate mass which is obtained from the top quark and the remaining b-jet in the event. The mass window obviously depends on the charged Higgs mass and is selected according to the obtained distribution shown on Fig. \ref{signals}. Table \ref{chmasses} shows the mass window cuts applied on the charged Higgs candidate mass distributions and the number of signal and background events remaining inside the mass window. The signal significance is then expressed in terms of $S/\sqrt{S+B}$ which is the number of signal events inside the mass window divided by the square root of the number of background events in the same window. Figure \ref{sigma} shows the signal significance as a function of the charged Higgs mass for different \tanb values used in the analysis.
\begin{table}
\begin{center}
\begin{tabular}{|l|c|c|c|c|c|c|c|c|}
\hline
Sample & \multicolumn{5}{|c|}{Signal} & $t\bar{t}$ & $WWb\bar{b}$ & QCD\\ 
\hline
$m_{H^{\pm}}$ [GeV]& 200 & 250 & 300 & 350 & 400 & - & - & - \\ 
\hline
6 jets eff. & 13 & 25 & 27 & 27 & 25 & 28 & 25 & 0.14\\ 
\hline
2 b-jets eff. & 69 & 72 & 69 & 67 & 65 & 73 & 61 & 0\\ 
\hline
Mass window eff. & 15 & 25 & 32 & 30 & 24 & 14 & 14 & 0\\ 
\hline
Total eff. & 1.3 & 4.5 & 6 & 5.4 & 4 & 3 & 2.2 & 0\\
\hline
\end{tabular}
\caption{Selection efficiencies of the signal and the background samples. \label{seleff}}
\end{center}
\end{table}

\begin{table}
\begin{center}
\begin{tabular}{|c|c|c|c|c|}
\hline
$m_{H^{\pm}}$ [GeV] & Mass Window [GeV] & S inside & $t\bar{t}$ inside & $WWb\bar{b}$ inside\\ 
\hline
200 &  170-200 & 1.8 & 0.89 & 0\\ 
\hline
250 & 220-250 & 11 & 78 & 26\\
\hline
300 & 250-300 & 16 & 163 & 46\\
\hline
350 & 300-350  & 14 & 116 & 39\\
\hline
400 & 350-400 & 6 & 38 & 20\\
\hline
\end{tabular}
\caption{The mass windows applied on the charged Higgs invariant mass distributions. The third, fourth and fifth columns show the remaining number of signal and background events inside the mass window. \label{chmasses}}
\end{center}
\end{table}
\section{Discussion on the Signal Observability}
\begin{figure}
\begin{center}
\includegraphics[width=0.7\textwidth]{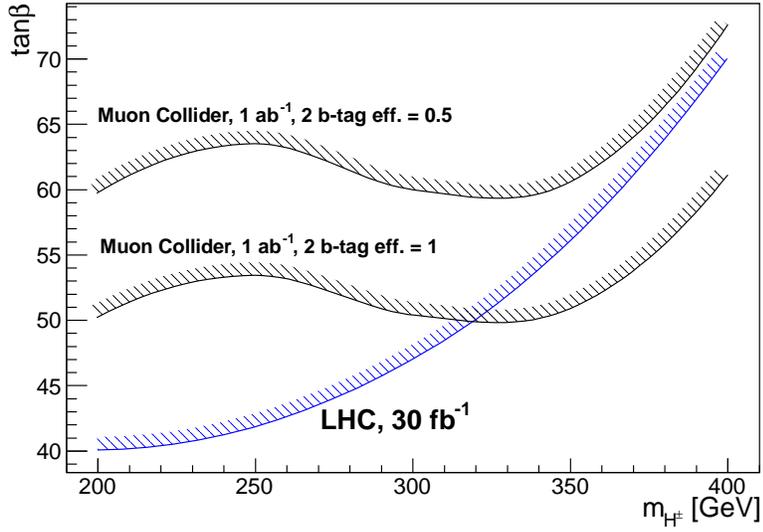}
\end{center}
\caption{The 5$\sigma$ contours of LHC and the muon collider for the observation of $W^{\pm}H^{\mp}$ channel.}
\label{contour}
\end{figure}

The resulting signal significances are not high enough for a positive perspective, however, some points should be mentioned in this direction. First the integrated luminosity used in the analysis is 10 times smaller than the corresponding value used for $e^{+}e^{-}$ linear collider analyses, i.e., 50 $fb^{-1}$ vs 500 $fb^{-1}$. It depends on the muon collider plan how to proceed with data taking when such a collider comes to operate. If the signal significance is calculated as $S/\sqrt{S+B}$, ignoring systematic uncertainties and their contribution in the significance formula, the significance grows like $\sqrt{L}$ where $L$ is the integrated luminosity. Therefore with \tanb = 50, roughly 1000 $fb^{-1}$ data is needed for the 5$\sigma$ discovery. If this amount of data becomes available, this channel could provide signals with the same size throughout the mass range $200 ~\textnormal{GeV}< m_{H^{\pm}} < 400$ GeV. Since the mass coverage is up to $\sqrt{s}-m_W \simeq 400$ GeV, such a low energy collider can perform more efficient than a linear $e^{+}e^{-}$ collider which has a charged Higgs mass reach up to $m_{H^{\pm}}<250$ GeV \cite{sch6} provided that the above amount of data is collected. Therefore we believe that the signal studied in this analysis can play an important role in the charged Higgs observation if enough amount of data is available. However at the time of obtaining these results, LHC would have already observed the charged Higgs signal if it exists. At the end it should be mentioned that the b-tagging algorithm used in this analysis was a simple one. The two b-jet tagging efficiency is expected to be about 50 $\%$ at the muon collider \cite{btag1,btag2}. Assuming this efficiency for signal and background samples, the signal significance is scaled by 70 $\%$. Therefore it was attempted to compare the muon collider observation potential with two assumptions of almost ideal b-tagging (eff. = 1) and two b-jet tagging efficiency of 50$\%$ with LHC results. Since the analysis uses $\mu=200$ GeV, we only found a CMS study with above $\mu$ value reported in \cite{cmscontour}. The result in terms of 5$\sigma$ contours is shown in Fig. \ref{contour}. The LHC result corresponds to an integrated luminosity of 30 $fb^{-1}$ while the muon collider result is shown with an integrated luminosity of 1 $ab^{-1}$.
\section{Conclusions}
A muon collider potential for the discovery of a charged Higgs boson in the associated production $H^{\pm}W^{\mp}$ was studied within the MSSM framework. It was shown that at 50 $fb^{-1}$ the signal significance in the fully hadronic final state is about 1$\sigma$ with \tanb = 50 for all charged Higgs masses in the range $200 < m_{H^{\pm}} < 400$ GeV. The signal observability depends on the amount of data taken by the muon collider, however the signal strength is the same for a charged Higgs mass in the studied range and can reach $5\sigma$ at about 1 $ab^{-1}$ integrated luminosity. Comparison with the LHC results shows that such a muon collider cannot add anything to the LHC findings.

\end{document}